# Recent results on the properties of two-phase argon avalanche detectors

A. Bondar, A. Buzulutskov*, A. Grebenuk, D. Pavlyuchenko, Y. Tikhonov

*Budker Institute of Nuclear Physics , 630090 Novosibirsk, Russia*

**Abstract**

The characteristic properties of two-phase Ar avalanche detectors, including those obtained with CsI photocathode, are further studied. Such detectors are relevant in the field of coherent neutrino-nucleus scattering and dark matter search experiments. The detectors investigated comprised a 1 cm thick liquid Ar layer followed by a triple-GEM multiplier. In these detectors, typical gains reaching $10^4$ were obtained with good reproducibility and a stable operation for at least one day was demonstrated. Amplitude and pulse-shape characteristics are presented under irradiation with X-rays, gamma-rays and neutrons from different radioactive sources. The detection of both primary scintillation and ionization signals at higher gains, at a deposited energy of 60 keV, has been demonstrated.
© 2008 Elsevier B.V. All rights reserved

*PACS:* 29.40.Cs; 34.80.My

*Keywords:* Two-phase avalanche detectors; Gas electron multipliers; Liquid argon; CsI photocathodes; Scintillations in noble liquids

## 1. Introduction

Two-phase avalanche detectors are referred to as those operated with noble liquids in an electron avalanching mode [1,2], using Gas Electron Multipliers (GEMs) [3] or thick GEMs (THGEMs) [4] operated in saturated vapor above the liquid phase. Such detectors have potential applications in rare-event experiments and in the medical imaging field: in particular, in coherent neutrino-nucleus scattering [5], dark matter search [6,7,8] and Positron Emission Tomography [2]. Their performances in Ar, Kr and Xe were described elsewhere [9,10,11,12,13,14,15]. Most promising results were obtained for the two-phase Ar avalanche detectors: using triple-GEM multipliers, gains reaching $10^4$ were obtained [12,13], permitting the operation in a single electron counting mode [13] and the detection of both primary scintillation and ionization signals using CsI photocathode [16]. High-gain operation of two-phase Ar avalanche detectors with a double-THGEM multiplier has also been recently demonstrated [17].

In this work we further study the performances of two-phase Ar avalanche detectors, including those with CsI photocathode. The detectors comprised a 1 cm thick liquid Ar layer followed by a triple-GEM multiplier. Some characteristic properties of the experimental setup are described. Gain, stability, amplitude and pulse-shape characteristics are presented under irradiation with X-rays, gamma-rays and neutrons from different radioactive sources. The detection of both primary scintillation and ionization signals at higher gains is demonstrated.

## 2. Experimental setup

A description of the experimental setup and procedures, with regard to the performance of two-phase avalanche detectors in Ar, Kr and Xe with a 3 and 10 mm thick liquid layer were presented in [10,11,12] and [13,17] respectively. Here we describe some details relevant to the detector performance in Ar with a 1 cm thick liquid layer, not reflected in previous works, including the measurement of the liquid layer thickness and the purification procedure.

A design of the cryogenic chamber is shown in Fig. 1. It had a volume of 2.5 l and comprised two 0.1 mm thick stainless steel windows to transmit the ionization radiation from the bottom, a cathode mesh at the chamber bottom immersed in the liquid and a triple-GEM assembly of an active area of 28×28 mm$^2$ placed in saturated vapor above the liquid. Typical distances were 12-13 mm between the cathode and the first GEM, 2-3 mm between the liquid surface and the first GEM and 2 mm between the GEMs. Note that at this configuration of the electrodes the electric drift field within the liquid is identical to the extraction field at the liquid-gas interface.

---
\* Corresponding author. Tel.: +7-383-3294833; fax: +7-383-3307163.
 *E-mail address*: a.f.buzulutskov@inp.nsk.su.

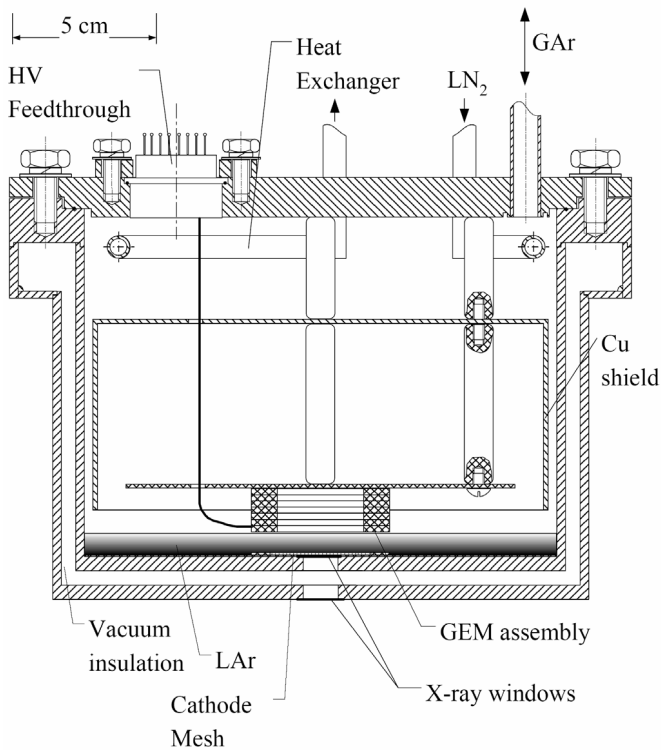

Fig. 1. Cryogenic chamber of a two-phase Ar avalanche detector.

The chamber was cooled from the top by liquid nitrogen using a heat exchanger mounted on the flange. The anode signals were read out from either the last electrode of the last

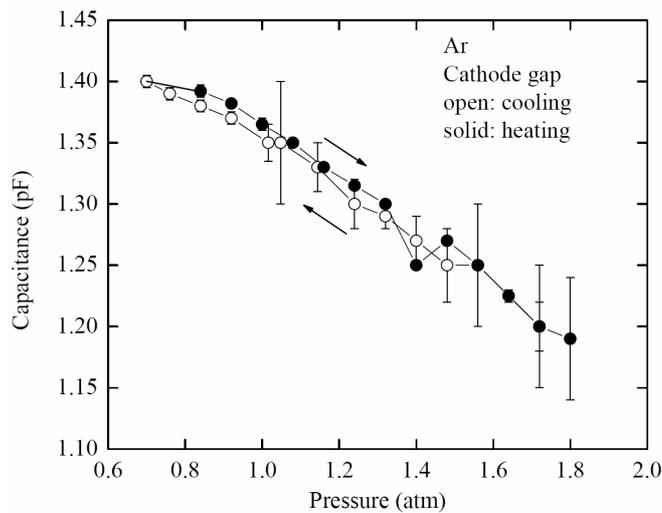

Fig. 2. Cathode gap capacitance as a function of the gas/vapor pressure during cooling and heating procedures in a two-phase Ar detector. The maximum capacitance corresponds to a liquid layer thickness of 1cm.

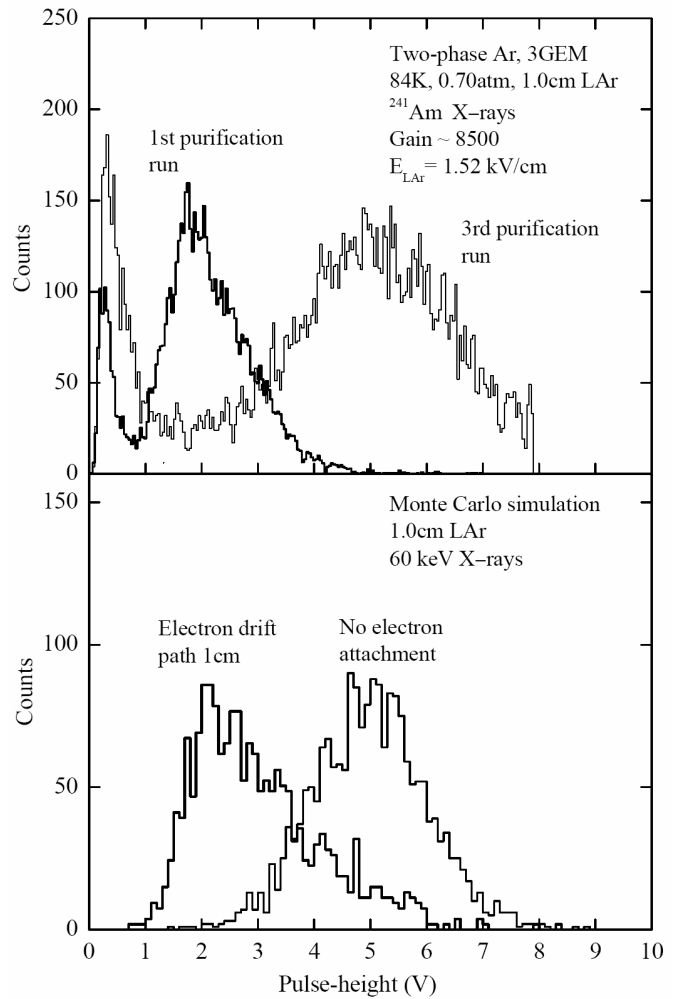

Fig. 3. Top: Pulse-height spectra of anode signals of a triple-GEM multiplier in two-phase Ar after the first and the third purification run. The signals are induced by 60 keV X-rays from $^{241}$Am source, at a gain of 8500 and drift field in the liquid of 1.52kV/cm Bottom: pulse-height spectra obtained by Monte-Carlo simulation for the electron drift path of 1 cm and with no electron attachment in the liquid.

GEM (3GEM operation mode) or the Printed Circuit Board (PCB) electrode behind the last GEM (3GEM+PCB mode). The amplifier shaping time was either 0.5 or 10 μs; the former was used for pulse-shape analysis and the latter for gain and amplitude spectrum measurements.

The liquid layer thickness at the chamber bottom was monitored by measuring the gap capacitance between the cathode and the first GEM, similarly to that described in [11,12]. Fig. 2 shows a dependence of the cathode gap capacitance on the gas/vapor pressure during cooling and heating procedures. At a temperature and pressure stabilization point, corresponding to 84 K and 0.70 atm, the liquid layer thickness was 10 mm. Note that capacitance fluctuations at this point were rather small; accordingly, the measurement accuracy of the liquid layer thickness here was better than 0.5 mm.

The Ar was purified using an Oxisorb filter placed between the bottle, supplying the gas to the system, and the cryogenic chamber. The purification run consisted of gas flowing through the filter during either cooling or heating procedures.

The Ar purity was monitored by measuring the position and the shape of the peak in pulse-height spectra of anode signals induced by 60 keV X-rays from $^{241}$Am radioactive source. In Fig. 2 the spectra obtained after the first and the third purification run are compared. One can see that after the first run the liquid Ar purity was not high enough, resulting in reduced amplitudes of the anode signals and asymmetrical spectrum shape. After the third purification run the anode amplitudes stopped increasing and the spectrum shape became symmetrical, indicating that the electron drift pass in the liquid before attachment exceeded the layer thickness, i.e. it was substantially larger than 1 cm. The latter corresponded to electron life times larger than 4 µs. Moreover, analyzing the pulse shape of anode signals observed at a low extraction field (see section 4), one may conclude that the electron life time in the liquid can be as large as 25 µs, after several purification runs. The spectra positions and shapes are well reproduced by Monte-Carlo simulation, accounting for X-ray absorption and electron attachment processes in the liquid (see Fig. 2): one can see that after the first purification run the spectrum shape corresponded to the electron drift path in the liquid of about 1 cm.

Other details of the experimental setup and procedures can be found elsewhere [10,11,12,13,17].

## 3. Gain, stability and amplitude characteristics

Gain characteristics of triple-GEM multipliers in two-phase Ar were studied earlier in [12,13]. Gain values as large as $10^4$ were reached there. In the present work we confirm the performance of two-phase Ar avalanche detectors at high gains; in addition, their reproducibility has been studied. In particular, in Fig. 4 the typical gain characteristics are compared: those for the same GEM assembly in the measurements separated in time by a two-week period, those for the same GEM assembly in a 3GEM and 3GEM+PCB mode and that for the other GEM assembly with a CsI photocathode deposited on the first GEM. One can see that gain characteristics are reproduced fairly well.

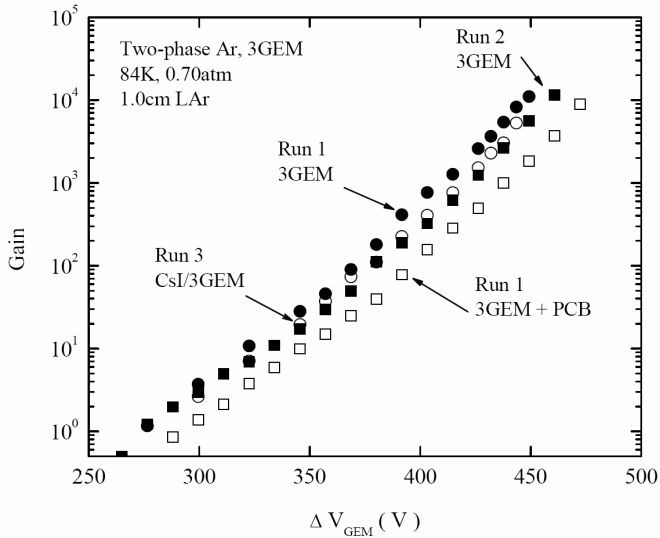

Fig. 4. Triple-GEM gain as a function of the voltage across each GEM in two-phase Ar obtained using pulsed X-rays. The typical gain characteristics are shown: for the same GEM assembly in 3GEM mode in the measurement runs separated by two-week period (Run 1 and Run 2, 3GEM), for the same GEM assembly in 3GEM and 3GEM+PCB mode (Run 1, 3GEM and 3GEM+PCB), for the other GEM assembly in 3GEM mode with CsI photocathode on the first GEM (Run 3, CsI/3GEM). The maximum gains are limited by discharges.

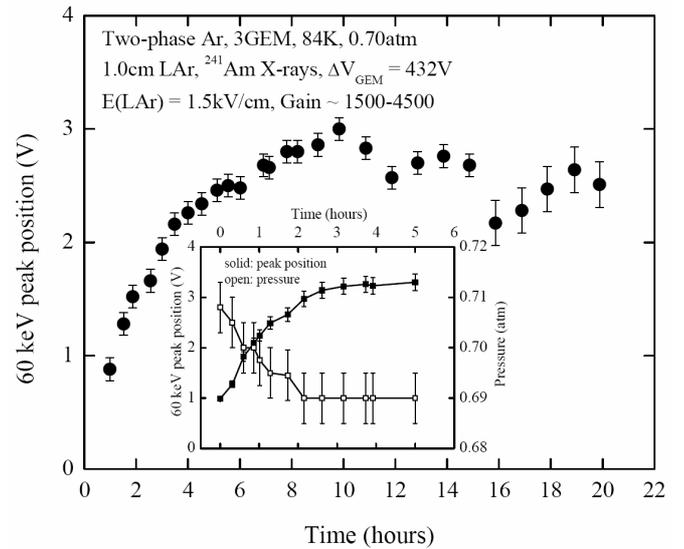

Fig. 5. Stability of operation of a two-phase Ar avalanche detector. A 60 keV peak position in the pulse-height spectrum of the triple-GEM induced by X-rays from $^{241}$Am source is shown as a function of time, at a gain of 1500-4500. In the inset both the peak position and the vapor pressure as a function of time are shown.

Fig. 5 illustrates the operation stability of the two-phase Ar avalanche detector. Compared to previous results [13], the gain was higher, reaching 4500, and the measurement period was increased from 5 to 20 hours. During this period the detector was constantly kept under voltage. One can see that the initial increase of the gain during first several hours is correlated to the decrease of the vapor pressure due to the temperature stabilization period. Taking this into account, one may conclude that two-phase Ar avalanche detectors can operate rather stably in a few-day scale, at gains of several thousands.

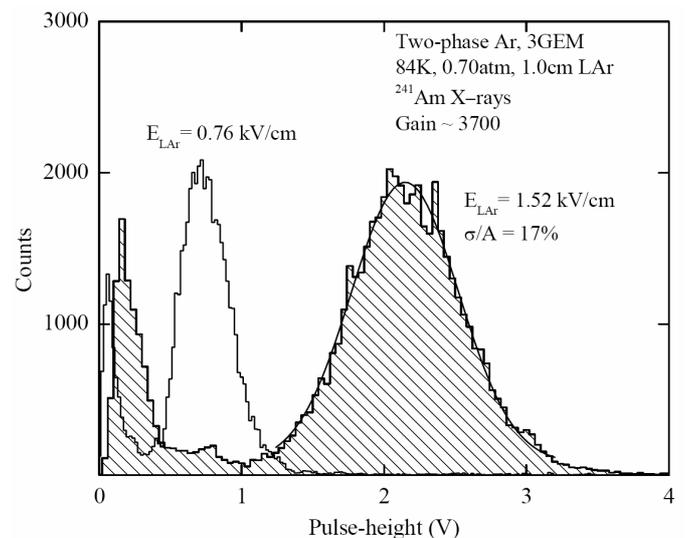

Fig. 6. Pulse-height spectra of anode signals of a triple-GEM multiplier in a two-phase Ar avalanche detector induced by $^{241}$Am X-rays at a gain of 3700 and different drift (extraction) fields in the liquid, of 1.52 and 0.76 kV/cm.

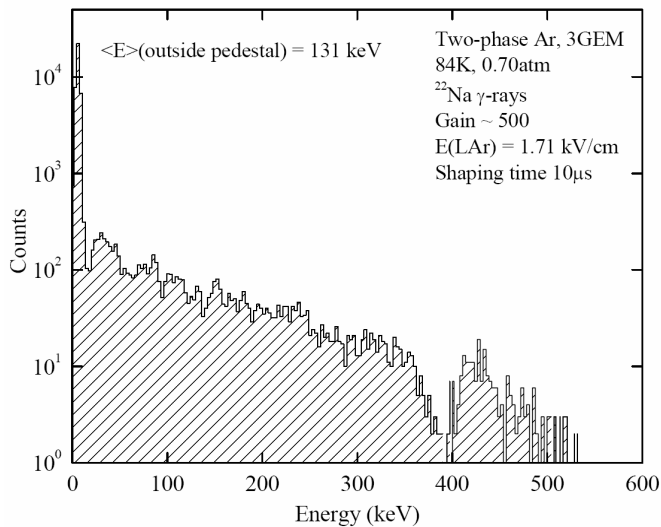

Fig. 7. Energy spectrum in a two-phase Ar avalanche detector induced by 511 keV gamma-rays from $^{22}$Na source at a gain of 500 and drift field in the liquid of 1.71 kV/cm.

A detector response to different radioactive sources is illustrated in Figs. 6-8. In Fig. 6 pulse-height spectra induced by X-rays from $^{241}$Am source are shown at different drift (extraction) fields in the liquid; the pulse-height resolution is the same as that reported earlier [12]: it amounts to 17% RMS at 60 keV. Fig. 7 and 8 show energy spectra induced by 511 keV gamma-rays from $^{22}$Na source and by neutrons and gamma-rays from $^{252}$Cf source, respectively. Here the 60 keV X-ray peak from $^{241}$Am source was used to calibrate the energy scales. The spectra in Figs. 7 and 8 were obtained in a triggered mode; the trigger was provided by one of the collinear 511 keV gamma-quanta and by gamma-quanta accompanying the nuclear fission, respectively, using a BGO scintillation counter. In Fig. 7 the Compton edge and the photoelectric absorption peak are distinctly seen. In Fig. 8 the bump between 3 and 15 keV is presumably due to neutron elastic scattering off nuclei (see next section).

## 4. Detection of weak signals

In coherent neutrino scattering and dark matter search experiments, the two-phase avalanche detectors should be able to detect weak signals induced by nuclear recoils with typical signal amplitudes of the order of 1-2 primary electrons and 1-50 keV of deposited energy, respectively. For the latter, the neutron elastic scattering off nuclei effectively imitates the nuclear recoil process.

In the current work we present preliminary results on the sensitivity of two-phase Ar avalanche detectors to nuclear recoils induced by neutron elastic scattering. These were obtained using a $^{252}$Cf spontaneous fission source, producing 4 neutrons and 20 gamma-quanta per fission. The neutrons have a Maxwellian energy distribution with a mean of 2.1 MeV [18], while 80% of gamma-rays have the energy below 1 MeV.

The appropriate energy spectrum obtained in a two-phase Ar avalanche detector in a triggered mode (see section 3) is shown in Fig. 8. The measurement was conducted with a 2.4 cm thick lead screen placed between the source and the cryogenic chamber, to reduce the gamma-ray background. In order to estimate the contribution of this background, the measurement with a $^{22}$Na source, providing 511 keV gamma-rays, was also done. Comparing these two spectra in Fig. 8, one may conclude that in the spectrum induced by $^{252}$Cf source the high-energy tail is due to the gamma-ray background, while the bump in the low-energy region, between 3 and 15 keV, might be attributed to neutron elastic scattering off nuclei. This conclusion is also supported by computer

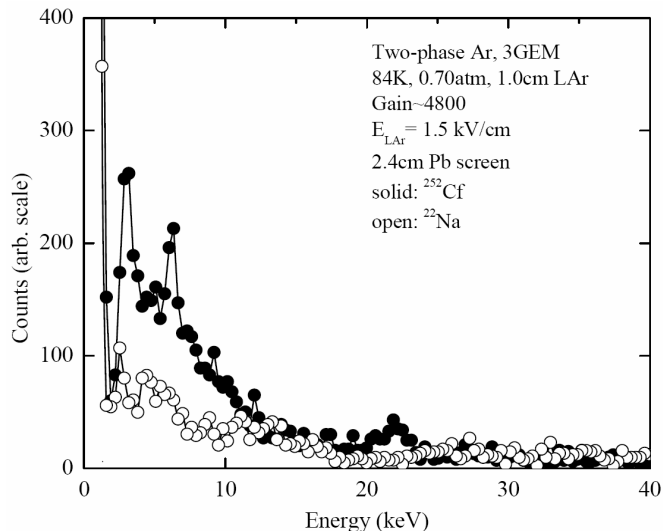

Fig. 8. Energy spectra in a two-phase Ar avalanche detector induced by neutrons and gamma-rays from $^{252}$Cf source and by gamma-rays from $^{22}$Na source, at a gain of 4800 and drift field in the liquid of 1.5kV/cm. There was a 2.4 cm thick lead screen between the source and the chamber.

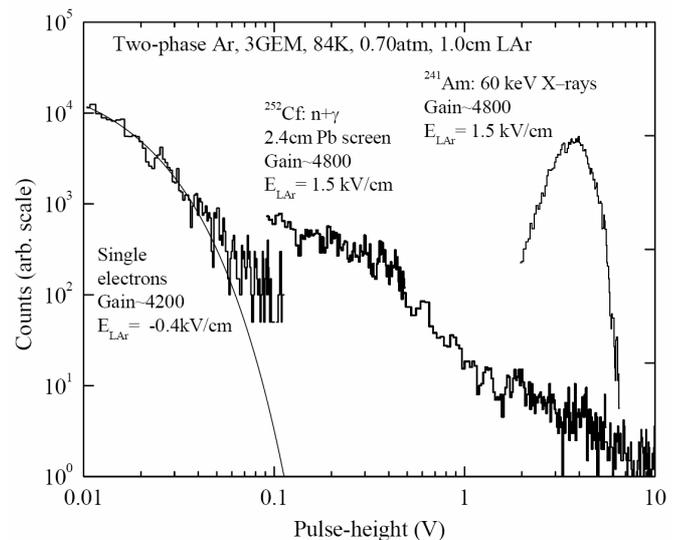

Fig. 9. Pulse-height spectra in two-phase Ar avalanche detectors induced by single electrons, neutrons and gamma-rays from $^{252}$Cf source and 60 keV X-rays from $^{241}$Am source, at gains in the range of 4000-5000. The single electron spectrum is fitted by an exponential function.

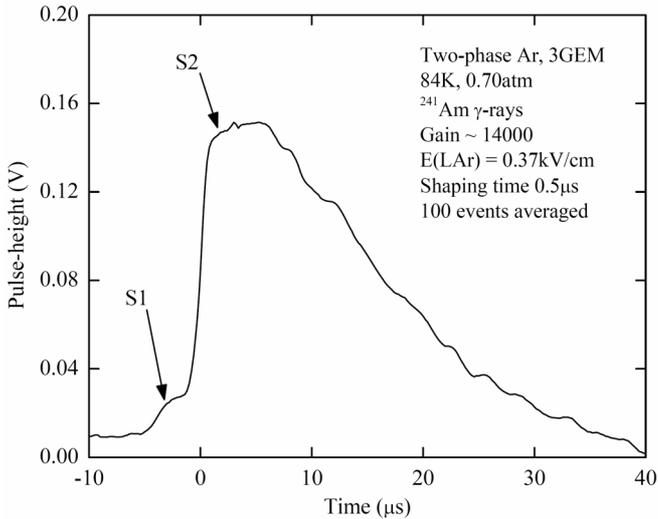

Fig. 10. Anode signals in a two-phase Ar avalanche detector with CsI photocathode induced by 60 keV X-rays from [241]Am source, at a drift field in the liquid of 0.37 kV/cm and gain 14000. The trigger threshold was 90 mV. The signals are averaged over 100 events.

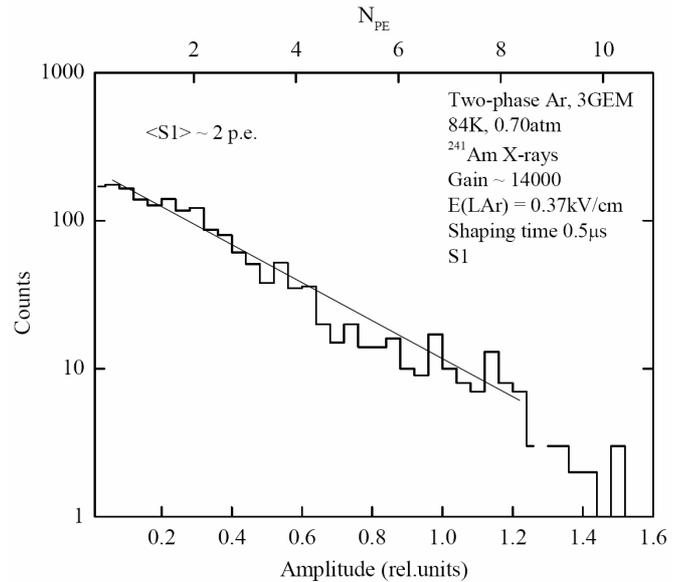

Fig. 11. Amplitude spectrum of primary scintillation (S1) signals in a two-phase Ar avalanche detector with CsI photocathode induced by 60 keV X-rays from [241]Am source at a gain of 14000 and drift field of 0.37 kV/cm. The top scale represents the number of photoelectrons prior to multiplication ($N_{PE}$).

simulation of the nuclear recoil spectrum, induced by neutron scattering, presented elsewhere [19].

The wide dynamical range of two-phase Ar avalanche detectors when detecting weak signals should be emphasized; it is illustrated in Fig. 9. Pulse-height spectra induced by single electrons (produced using a technique described in [13]), neutrons and gamma-rays from [252]Cf source and 60 keV X-rays from [241]Am source are shown in the figure in double-logarithmic scale, at gains in the range of 4000-5000.

The detection of both primary scintillation (S1) and ionization (S2) signals in two-phase detectors is of particular importance for effective background rejection in dark matter search experiments [6,7,8]. This technique has been recently demonstrated in two-phase Ar avalanche detectors based on triple-GEM multipliers, using a CsI photocathode deposited on the first electrode of the first GEM [16]. In that work the signals were produced by beta-particles and the energy deposition in the liquid was 600 keV on the average. In the present work we further demonstrate this technique at a lower deposited energy, of 60 keV. This was achieved by operation of the triple-GEM multiplier at higher than typical gains, namely at 14000.

Fig. 10 shows anode signals in a two-phase Ar avalanche detector with CsI photocathode induced by 60 keV X-rays from [241]Am source at a drift field in the liquid of 0.37 kV/cm. One can see that in addition to the ionization signal (S2) the primary scintillation signal (S1) was observed, preceding S2. At such a low electric field, needed to separate S1 from S2, the ionization signal was rather slow due to delayed electron emission through the liquid-gas interface [17]; the decay time of the signal was as large as 25 μs. This particular time is most probably limited by the electron life-time in the liquid, i.e. it is defined by the liquid purity (see section 2).

Following the procedure described in [16], the amplitude spectrum of the primary scintillation (S1) signals was obtained; it is shown in Fig. 11. One can see that this spectrum is described by an exponential function, which is typical for gas avalanche detectors operating in a few electrons counting mode. Taking into account the average spectrum charge and the detector gain, one can express the S1 amplitude in the number of photoelectrons prior to multiplication: it amounts to ~2 photoelectrons. It should be remarked that just the operation at gains exceeding $10^4$ allowed one to effectively detect such a weak signal.

## 5. Conclusions

The performances of two-phase Ar avalanche detectors were further studied. The detectors comprised a 1 cm thick liquid Ar layer followed by a triple-GEM multiplier, sometimes coated with a CsI photocathode.

Some characteristics properties of the experimental setup were described, including the purification and the liquid level measurement procedures. The detector operation at typical gains reaching $10^4$, with good reproducibility, was confirmed. The stable detector operation during 20 hours at a gain of several thousands was demonstrated.

X-rays from [241]Am source, gamma-rays from [22]Na source and neutrons and gamma-rays from [252]Cf source were recorded in the two-phase Ar avalanche detector with a reasonable energy resolution (17% RMS at 60 keV). Preliminary results on the detector sensitivity to nuclear recoils induced by neutron elastic scattering were presented.

The detection of both primary scintillation and ionization signals in the two-phase Ar avalanche detector with CsI photocathode was demonstrated at a deposited energy of 60 keV.

The results obtained are relevant in the field of low-background detectors sensitive to nuclear recoils, such as those for coherent neutrino-nucleus scattering and dark matter search experiments. Further studies of this technique are in progress.

## References

1. B.A. Dolgoshein et al., Phys. Elem. Part. Atom. Nucl. 4 (1973) 167 (in Russian).
2. A. Buzulutskov, Instr. Exp. Tech. 50 (2007) 287, and references therein.
3. F. Sauli, Nucl. Instr. and Meth. A 386 (1997) 531.
4. C. Shalem et al., Nucl. Instrum. Meth. A 558 (2006) 475, and references therein.
5. C. Hagmann, A. Bernstein, IEEE Trans. Nucl. Sci. NS-51 (2004) 2151.
6. E. Aprile et al., J. Phys. Conf. Series 39 (2006) 107.
7. P. Benetti et al., Astroparticle Physics 28 (2008) 495.
8. A. Rubbia, J. Phys. : Conf. Ser. 39 (2006) 129.
9. A. Buzulutskov et al., IEEE Trans. Nucl. Sci. NS-50 (2003) 2491.
10. A. Bondar et al., Nucl. Instrum. Meth. A 524 (2004) 130.
11. A. Bondar et al., Nucl. Instrum. Meth. A 548 (2005) 439.
12. A. Bondar et al., Nucl. Instrum. Meth. A 556 (2006) 273.
13. A. Bondar et al., Nucl. Instrum. Meth. A 574 (2007) 493.
14. V. Solovov et al., Nucl. Instrum. Meth. A 580 (2007) 331.
15. V. Peskov et al., IEEE Trans. Nucl. Sci. NS-54 (2007) 1784.
16. A. Bondar et al., Nucl. Instr. and Meth. A 581 (2007) 241.
17. A. Bondar et al., Eprint arXiv:0805.2018, 2008.
18. G.V. Meadows, Phys. Rev. 157 (1967) 1076.
19. P. Mijakowski, Acta Phys. Polonica B 37 (2006) 2179.